\documentclass[lettersize,journal]{IEEEtran}
\usepackage[table,xcdraw]{xcolor}
\usepackage{xcolor,soul,framed} 
\colorlet{shadecolor}{yellow}
\usepackage[pdftex]{graphicx}
\graphicspath{{../picture/}{../jpeg/}}
\DeclareGraphicsExtensions{.pdf,.jpeg,.png}
\usepackage[cmex10]{amsmath}
\usepackage{amssymb}
\usepackage{array}
\usepackage{mdwmath}
\usepackage{mdwtab}
\usepackage{eqparbox}
\usepackage{url}
\usepackage{makecell}
\usepackage{multirow}
\usepackage{longtable}
\usepackage{cite} 
\usepackage{rotating}

\usepackage{enumerate}
\usepackage{threeparttable} 
\usepackage{color}
\usepackage{tabularray}
\usepackage{CJK}
\usepackage[top=2cm, bottom=2cm, left=2cm, right=2cm]{geometry}
\usepackage[ruled,vlined,linesnumbered]{algorithm2e}
\usepackage{algpseudocode}
\usepackage{soul}
\usepackage{booktabs}
\usepackage{subfigure} 
\usepackage[colorlinks=true,linkcolor=black,citecolor=blue,urlcolor=blue,]{hyperref}
\usepackage[doipre={doi:~}]{uri}

\soulregister\cite7
\soulregister\ref7

\begin{document}

\title{A Spatio-Temporal Approach with Self-Corrective Causal Inference for Flight Delay Prediction}

\author{
\IEEEauthorblockN{Qihui Zhu, Shenwen Chen, Tong Guo, Yisheng Lv,~\IEEEmembership{Senior Member,~IEEE}, Wenbo Du,~\IEEEmembership{Member,~IEEE}}

\thanks{Manuscript received November 30, 2023. \textit{(Corresponding author: Wenbo Du, e-mail:wenbodu@buaa.edu.cn) }}
\thanks{Qihui Zhu, Shenwen Chen, Tong Guo and Wenbo Du are with the State Key Laboratory of CNS/ATM, the School of Electronic and Information Engineering, Beihang University, Beijing 100191, China.}
\thanks{Yisheng Lv is with the State Key Laboratory of Multimodal Artificial Intelligence Systems, Institute of Automation, Chinese Academy of Sciences, Beijing 100190, China, also with the Shandong Key Laboratory of Smart Transportation (Preparation), Shandong Jiaotong University, Jinan 250353, China, and also with the School of Artificial Intelligence, University of the Chinese Academy of Sciences, Beijing 100049, China.}}

\markboth{IEEE TRANSACTIONS ON INTELLIGENT TRANSPORTATION SYSTEMS,~Vol.~X, No.~X, November~2023}%
{Zhu \MakeLowercase{\textit{et al.}}: A Spatio-Temporal Approach with Self-Corrective Causal Inference for Flight Delay Prediction}


\maketitle

\begin{abstract}
Accurate flight delay prediction is crucial for the secure and effective operation of the air traffic system. Recent advances in modeling inter-airport relationships present a promising approach for investigating flight delay prediction from the multi-airport scenario. However, the previous prediction works only accounted for the simplistic relationships such as traffic flow or geographical distance, overlooking the intricate interactions among airports and thus proving inadequate. In this paper, we leverage casual inference to precisely model inter-airport relationships and propose a self-corrective spatio-temporal graph neural network (named CausalNet) for flight delay prediction. Specifically, Granger causality inference coupled with a self-correction module is designed to construct causality graphs among airports and dynamically modify them based on the current airport’s delays. Additionally, the features of the causality graphs are adaptively extracted and utilized to address the heterogeneity of airports. Extensive experiments are conducted on the real data of top-74 busiest airports in China. The results show that CausalNet is superior to baselines. Ablation studies emphasize the power of the proposed self-correction causality graph and the graph feature extraction module. All of these prove the effectiveness of the proposed methodology. 
\end{abstract}


\section{Introduction}
\IEEEPARstart{I}{n} recent years, despite the impact of the COVID-19 epidemic, the global air transport industry has continued to develop rapidly.
However, this surge in demand for air travel, paired with limited capacity, has intensified the issues of aviation network congestion and flight delays. Eurocontrol's 2022 report notes that while traffic flows has reached 83\% of their 2019 levels in 2022, flights are facing significant punctuality challenges. In the summer of 2022, airlines recorded average flight schedule delays of 20.4 minutes per aircraft, an increase of 41.67\% from 2019\cite{eurocontrol2022}. 
These frequent and inevitable delays not only strain airport resource reallocations and disrupt flight schedules, but lead to significant economic repercussions. 
By 2022, flight disruptions will cost the United States and Europe \$34 billion and \$32 billion respectively, an increase of 11\% from 2019, and a total of more than 650 million hours of passenger time will be wasted\cite{Airhelp}.

The ability to predict flight delays offers an opportunity for air traffic system to devise timely and effective strategies, thereby mitigating economic impacts.However, predicting delays comes with its own set of challenges
\cite{DU_2,DU_1,WU_propagation,CAIgeo} since it is influenced by several interrelated factors, including airport operational capacity, geographical and meteorological conditions, and air traffic control \cite{li2023multi,LAMBELHO2020101737,delay_reason2}.
Initial efforts for flight delay prediction primarily relied on simulations or probabilistic statistics related to air traffic operations~\cite{wesonga2012parameterized,Barnhart2014ModelingPT,Gur2014}. While these techniques offered crucial insights into airport operations, they often demanded a more extensive set of parameters or assumptions. The surge in available air traffic data ushered in the use of data-driven machine learning methods to predict the flight delays. 
Starting with the employment of simple statistical regression models and classical machine learning methods\cite{delay_reason2,hao2014newYork}, the evolution toward more sophisticated deep learning models, such as Recurrent Neural Networks (RNNs), Long Short-Term Memory units (LSTMs), and Gated Recurrent Units (GRUs), has markedly enhanced prediction accuracy \cite{Ai2019ADL,gui2019flight}.
Although the majority of these methods concentrate on modeling flight delays at a single airport, it is crucial to note that delays across various airports may exhibit spatial correlations due to the transmission of traffic among them.

As an emerging technique, Spatio-Temporal Graph Neural Networks (STGNNs) effectively integrate spatial relationships with temporal dynamics, providing precise predictions and showcasing versatility across multiple research domains \cite{TGCN,ITSM_new1,ITSM_new3,STGCN}. This innovative approach further sparked interest in examining how flight delays influence each other through time series predictions \cite{shenwen,BAO2021103323,Cai}. 
The results indicate that STGNNs methods, which take into account both inter-airport delay relationships and temporal dependencies, outperform models that focus only on temporal dependencies.
Most existing STGNNs-based methods utilized geographic distance or air traffic volume between airports to characterize the delay propagation in space dimension. 
However, the delays propagation between airports is affected by multiple factors. Due to resource links such as aircraft rotations, passenger links, and security checks, delayed flights from upstream airports landing at downstream airports will occupy resources, affect the operational efficiency of downstream airports, and create a delay propagation effect, making it very difficult to accurately characterize the delay propagation of airports. Simply utilizing the geographic distance or traffic volume may not be sufficient to capture such complicated relations.

The \textit{causal inference} is an effective method to model the delay propagation between airports \cite{DU2018,Guo2022DetectingDP}. Given two delay sequences of two airports, the causal inference  adopts the Granger causality \cite{granger_2001} to quantify the inter-airport delay correlation.
Inspired by this, it is possible to leverage the causal inference to enhance the performance of STGNNs on flight delay prediction by providing a casual graph representing the more accurate spatial correlation of airports.

The accuracy of causal inference depends on the stationarity of the data~\cite{nature_causal,causal_nonstationary_zhou,causal_nonstationary_xu}. However, in the air traffic system, flights operate in a cyclical manner with complex trends within each cycle, resulting in a strong non-stationarity in the delay sequences~\cite{yinguo_Zeng_2022,yinguo_Liyue_2023}. Differencing multiple times is a common method to eliminate the non-stationarity of a series, but this approach may compromise the temporal information inherent in the series, leading to biases in the calculated causal values~\cite{Moraffah_2021,Granger_pro1}. Therefore, directly adopting the existing causal inference for flight delay prediction may become less effective. 

Furthermore, due to varying capacities and operational management abilities, different airports are affected by delay propagation to different extents~\cite{csw1,shenwen,lv1}. Traditional GCN modules are incapable of extracting the heterogeneity of airports from the causal graph. Moreover, simply integrating them into existing STGNNs may not effectively represent the spatial correlations of flight delays. Thus, addressing the characteristics of each airport is a crucial step in accurately extracting spatial correlation information.

To address both the above-mentioned challenges, we propose a self-corrective spatio-temporal graph neural network with causal inference, termed \textbf{CausalNet}, for precise flight delay prediction.  
Specifically, unlike the existing STGNNs-based approaches, to accurately characterize the delay propagation in space dimension, CausalNet adopts Granger causality inference to construct a causal graph among multiple airports. 
To alleviate the information loss when constructing causal graphs, a \textit{self-causal correction module} with trainable parameters is further devised to adaptively modify the elements of the causal graph.
Furthermore, to extract more accurate spatial correlation information, an information extraction method considering heterogeneity based on graph convolution is proposed in \textit{spatial dependence modeling}. Moreover, \textit{long-gate recurrent units(LGRUs)} are leveraged to encapsulate temporal dependencies. 
Comprehensive experimental results, utilizing real-world air traffic data, reveal that CausalNet outperforms state-of-the-art approaches across varied prediction horizons in terms of prediction accuracy. Further, the self-corrective causal inference model proves adept at providing a more accurate depiction of delay propagation effects.

In a nutshell, the contributions of the paper are:
\begin{enumerate}
    \item A spatio-temporal graph neural network with self-corrective causal inference is proposed for the flight delay problem.
    \item Verifying the effectiveness and superiority of the developed approach across real-world air traffic data.
    \item Our results provide insightful guidance for air traffic management that smaller airports are more susceptible to the influence of other airports.
\end{enumerate} 

The arrangement of  this paper is as follows: Section 
$\text {II}$ reviews the existing literature on the prediction of flight delay and provide the problem formulation. Section $\text {III}$ elaborates the model we constructed. The experimental results on the real dataset of Chinese airports are given in Section $\text {IV}$.

\section{Background}
\subsection{Related work}
\subsubsection{\textbf{Simulation and probability statistics methods}}
Initial efforts for flight delay prediction primarily relied on simulations or probabilistic statistics related to air traffic operations. The simulation method used by Wu et al~\cite{wu2018modeling} mimics real-life scenarios to predict delays. Their model integrated a queuing engine, link transmission model, and delay propagation model while considering airport and airspace capacity. However, the required simplifications in simulations limited their real-world applicability and increased computational demands.

Probability and statistics methods use mathematical models to estimate the distribution function or probability of flight delays. Tu et al.~\cite{tu2008estimating}  proposed a nonparametric, mixed-distribution model for takeoff delays, optimized using expectation maximization and genetic algorithms. Abdel-Aty et al.~\cite{delay_reason1} employed a two-stage analysis process, with the first stage using frequency analysis methods to detect periodicity in delay data and the second stage using statistical methods to identify factors associated with delays. They demonstrated their analysis on arrival delay data for flights at Orlando International Airport in 2002–2003 and found seasonal, monthly, and daily delay patterns. While these techniques offered crucial insights into airport operations, they often demanded a more extensive set of parameters or assumptions.


\subsubsection{\textbf{Statistical regression models}}
With the availability of massive aviation data, data-driven regression methods have attracted attention due to their ability to integrate various features influencing flight delays, fitting delay data accordingly.
Lu et al.~\cite{hao2014newYork} used a simultaneous equation regression model to consider the interaction between delays at New York airport and delays at other airports, and model parameters were estimated using a substantial dataset to predict the impact of delays.  
Guvercin~\cite{guvercin2020forecasting} implemented a two-step approach for predicting airport delays. Initially, the airport network is clustered; subsequently, a REG-ARIMA model—which amalgamates regression and Autoregressive Integrated Moving Average (ARIMA) models—is constructed. Despite their interpretability, regression methods necessitate considerable data preprocessing and exhibit limited capacity to accommodate nonlinearity.

\subsubsection{\textbf{Classical machine learning methods}}
With the advent of machine learning, the capability of classical machine learning methods to fit nonlinear regression equations has advanced, leading to enhanced prediction accuracy. Kalliguddi~\cite{kalliguddi2017predictive} utilized cleaned and estimated 2016 U.S. domestic flight data, employing techniques like decision trees and random forests for predicting and analyzing flight delays. Chen~\cite{chen2019chained}applied a multi-label random forest classification methodology to forecast delays for individual flights throughout their scheduled trip sequences. Yu et al.~\cite{yu2019flight} leveraged deep belief network methods in conjunction with support vector regression to mine delay internal patterns. Rodríguez-Sanz~\cite{rodriguez2019assessment} proposed a probabilistic model based on Bayesian networks and Markov processes, which can consider uncertain factors such as weather and airport capacity that affect flight operations and arrivals. 

\subsubsection{\textbf{Deep learning time series models}}
Over the past decade, recurrent neural networks (RNNs)~\cite{RNN}, long short-term memory units (LSTMs)~\cite{LSTM}, and gated recurrent units (GRUs)~\cite{GRU} have demonstrated remarkable capabilities in capturing temporal dependencies, inspiring numerous studies to employ time series models for delay prediction. Gui et al.\cite{gui2019flight}assessed LSTMs on various flight delay tasks, discovering that LSTMs effectively capture flight delays' time dependencies across multiple tasks. Kim~\cite{Kim2016ADL}introduced a two-stage model utilizing deep RNNs and neural networks(NNs) for predicting daily delay status and individual flight delays, respectively. Additionally, Wei et al.~\cite{Weiaerospace10070580}implemented a Bidirectional LSTM with an attention mechanism to identify key delay-influencing features. During feature extraction, Principal Component Analysis was employed to isolate crucial features, while using the correlation coefficient method to measure the relevance between each feature and the target variable, and selecting high-correlation features as model inputs.

\subsubsection{\textbf{Spatio-Temporal graph neural networks}}
Recently, the Graph Neural Network(GNN)~\cite{GNN2008}, which can efficiently extract spatial information, has attracted a lot of interest, and Graph Convolutional Neural Network (GCN)~\cite{GCN} has been further designed to fully exploit topological features. Since ground traffic has a relatively clear graph structure, STGNNs for spatial and temporal dependency capture has made a huge breakthrough in ground traffic delay prediction~\cite{lv2,MTGNN,STGCN}. 

This has successfully motivated research on flight delays using STGNNs. With the Air Transport System (ATS) being highly dynamic and interlinked, methods that take into account both inter-airport delay relationships and temporal dependencies outperform models only focusing on temporal dependencies and enhance prediction accuracy significantly.
Zeng et al.~\cite{Zeng}integrated traffic flow and geographic distance to depict delay relationships between airports. They developed a delay prediction framework named DGLSTM by constructing a diffusion convolution kernel within STGNNs to capture delay propagation characteristics.
Bao et al.~\cite{BAO2021103323} introduced AG2S-Net, a deep learning framework that merges attention mechanisms, GCN, and LSTM to predict airport delays, incorporating various inputs like weather variables and airline features.
Cai et al.~\cite{Cai} used inter-airport flow to represent delay correlations, employing Markov-based time convolution blocks to extract time-varying flight delay patterns through graph snapshots in their MSTAGCN model.
Zheng et al.~\cite{ZHENG2024new} introduced an external impact modeling module to consider the impact of weather on flight delay patterns, and developed a spatiotemporal gated multi-attention graph network to predict airport delays, improving the prediction accuracy.
Most existing STGNN-based methods primarily use geographic distance or air traffic volume to represent delay propagation between airports. However, this simplification is not sufficient to capture the complex relationships involved in delay propagation.

The causal inference is an effective method to model the delay propagation between airport. Du et al.~\cite{DU2018} developed a delay causality network based on the Granger causality test and investigated the impacts of delay propagation in an airport network. Zeng et al.~\cite{yinguo_Zeng_2022} proposed an extreme event measurement mechanism to identify extreme events at different airports and then combine it with the cause-and-effect diagram discovery algorithm PCMCI which can find causality at a fine granularity. Guo et al.~\cite{guvercin2020forecasting} applied the convergent cross mapping to quantify the delay causality among airports and developed a delay causality network to understand the delay propagation patterns in a regional airport network. Wang et al.~\cite{WANGyj}extracted statistically significant lags between airport delay time series based on the piecewise cross-correlation values. Causal methods are often used to calculate the correlation strength of delay propagation. However, the non-stationarity of the aviation delay time series can easily lead to causal calculation bias. Therefore, directly using the existing causal effects to predict flight delays may reduce the effect.

Inspired by leveraging inter-flight causal inference to provide a more accurate characterization of the spatial correlation and then enhance the performance of STGNNs, we propose \textbf{CausalNet}, a self-corrective spatio-temporal graph neural network incorporating causal inference for precise flight delay prediction. 
CausalNet uses Granger causality inference to construct causal correlation graphs between multiple airports, and further designs a \textit{self-causal correction module} to alleviate the information loss when constructing causal graphs. Furthermore, to extract more accurate spatial correlation information, 
an information extraction method considering heterogeneity based on graph convolution is proposed in \textit{spatial dependence modeling}. Moreover, \textit{long-gate recurrent units(LGRUs)} are leveraged to encapsulate temporal dependencies. Extensive experiments with real-world air traffic data demonstrate that CausalNet surpasses existing state-of-the-art methods in prediction accuracy across various horizons.

\subsection{Problem Formulation}
The air transportation system at time step $t$ is defined as a weighted graph $G^t=\left(V, E^t, A^t\right)$, where $V$ is a set of nodes with size $|V|=N$ which represents the number of airports in the system. $E^t$ denotes the set of edges in the graph indicating the relationship among nodes in $V$ at time step $t$ and $A^t \in R^{N \times N}$ is the weighted adjacency matrix which represents the strength of the relationships. Most existing works establish the graph using geographic distance or air traffic flow.

Consistent with previous delay prediction work, we divided daily delay data into 24 time slices with an interval of 1 hour, and the average delay time of each airport under each time slice was used as the delay status.
$Y^t=\left\{y_1^t, y_2^t, \ldots, y_N^t\right\} \in R^N$ is employed to denote an observation vector of $N$ airports at time $t$, of which each element $y_i^t$ records the delay state at a single airport $i$ during $(t-1, t)$.   
In computing the average flight delay time at airport $i$, it is imperative to consider not only the actual flight delay time but also flight cancellations, as they serve as crucial indicators for assessing flight punctuality. 
Therefore, according to the regulations of the US Federal Aviation Administration (FAA), the European Air Safety Organization (Eurocontrol) Network Management Center (NM) Operation Center and the Civil Aviation Administration of China (CAAC), $\rho$ is used to represent the equivalent delay time of flight cancellation. 
The average flight delay time $y_i^{t}$ is represented by a weighted value that combines the actual flight delay time and the equivalent cancellation delay time $\rho$~\cite{wu2018modeling}: 
\begin{equation}
y_i^{t}=\frac{m_i^{(t)}+\rho * c_i^{(t)}}{a_i^{(t)}}
\end{equation}
where $m_i^{(t)}$ reveals the total delay of departure flights at an airport $i$ during $(t, t-1)$; $c_i^{(t)}$ represents the number of cancelled flights, and $a_i^{(t)}$ represents the total number of scheduled departure flights at the airport $i$ during $(t, t-1)$. $\rho = 180$ minutes represents the equivalent delay time of a cancellation. 

Based on the above description, the delay prediction problem can be transformed into the following time series analysis task: in a specific time step $t$, use the current and past $r$ steps of features of each airport to predict the delay time of each airport in the future $m$ time steps:
\begin{equation}
\begin{gathered}
\hat{Y}^{t+1}, \hat{Y}^{t+2}, \ldots, \hat{Y}^{t+m}=\mathcal{F}\left(X^{t-r}, X^{t-(r-1)}, \ldots, X^t ; \theta\right) \\
\min _\theta \mathcal{L}(\theta)=\sum_{t+1}^{t+m} L\left(\hat{Y}^t(\theta), Y^t\right)
\end{gathered}
\end{equation}
In equation (2), $\hat{Y}^{t+1}, \hat{Y}^{t+2}, \ldots, \hat{Y}^{t+m}$ is the predicted value of $Y^{t+1}, Y^{t+2}, \ldots, {Y}^{t+m}$ in the future time period, calculated by the function $\mathcal{F}$, which takes as input the feature $X^{t-r}, X^{t-(r-1)}, \ldots, X^t$ of the previous time period and the trainable parameter vector $\theta$. $X^t=\left\{X_1^t, X_2^t, \ldots, X_N^t\right\} \in R^{N \times D}$ is employed to denote the features of $N$ airports, where $D$ is feature dimension. The optimization objective is to minimize the total loss function $\mathcal{L}(\theta)$, defined as the sum of scalar losses $L$ over time period. The loss function $L(\cdot)$ measures the difference between the predicted and observed values for each time period. The optimal parameter vector is obtained by solving the optimization problem.

\section{The proposed Method}
\subsection{Overview}
In this section, we proposed a self-corrective spatio-temporal graph neural network with causal inference, termed CausalNet, for precise flight delay prediction. 

As shown in Fig. 1, in the input layer, besides the geographic graph $A$ which is commonly used in the research of air traffic systems, we generate a series of delay causality graphs $C^{i},{i\in (t-r, \dots, t)}$ using the \textit{granger causal graph construction module} based on the historical flight delay data. 
In the traning layer, we designed an Encoder-Decoder framework to sequentially input delay features at different moments. In each encoding process, to obtain a more accurate causal relationship, we input the causality graph of the current moment $i$ into the \textit{self-causal correction module}, and aggregate the delay features $X^{i}$ and the hidden layer of the previous moment status $H^{i-1}$. The output correction mask $CM^i$ is linearly weighted to obtain self-corrected causal graph ${CA^i}$. 

Since different airports are affected to varying degrees by the propagation effect, to extract more accurate spatial correlation information, an information extraction method considering heterogeneity based on graph convolution is proposed in \textit{spatial dependence modeling}.  The normalized ${CA^t}$ and $A$ are input into GCN respectively to aggregate node information and extract spatial correlation, and the output ${FC^t}$ and ${FA^t}$ are adaptively weighted. The whole process loops $K$ stack updates, which is called $K$-hop GCN.

Subsequently, we use the \textit{long-gate recurrent units(LGRUs)} to retain long-term memory and extract time dependencies, and output the hidden layer state $H^{i}$ at current time. After $r$ Encoders and $m$ Decoders, the final output is converted through a linear layer to obtain the delay prediction of $m$ moments in the future. It should be noted that the delay feature $X$ becomes an all-0 vector of the same size at the Decoders.
\begin{figure*}[t]
  \centering
  \includegraphics[width=7in]{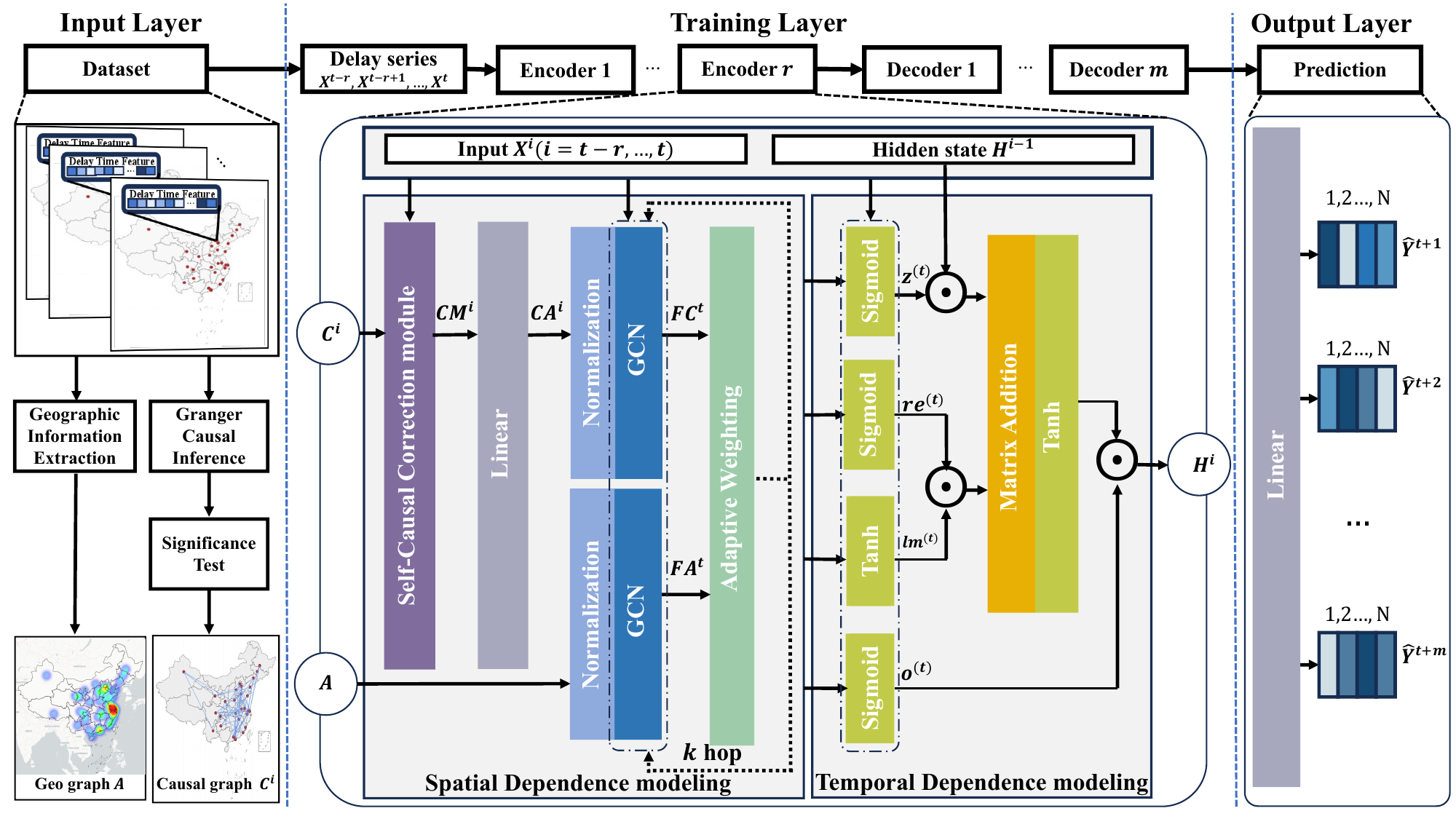}
\caption{The architecture of the proposed self-corrective spatio-temporal graph neural network.}
\label{fig_1}
\end{figure*}

\subsection{Granger Causal Graph Construction Module} 
The delay relationship within airports is a typical non-Euclidean structure, which can be represented by the weighted graph described in Section $\text {II}$. 
Inter-airport connections are affected by multiple factors. Due to resource links such as aircraft rotations, passenger links, and security checks, delayed flights from upstream airports landing at downstream airports will occupy resources, affect the operational efficiency of downstream airports, and create a delay propagation effect, making it very difficult to accurately characterize the delay propagation of airports. Simply utilizing the geographic distance or traffic volume are not be sufficient to represent such complicated relationships. 

In recent years, causal inference has been successful in mining delay propagation patterns in ATS. Existing work pioneered the discovery that the essence of delay propagation is the causal relationship of delayed flights.~\cite{yinguo_Zeng_2022,yinguo_Liyue_2023} Given delay sequences from two airports, causal inference can quantify inter-airport delay correlations.~\cite{Guo2022DetectingDP} Inspired by this,  Granger causality test is applied to construct a delay causality graph among airports.~\cite{DU2018}

In the air traffic system, flight operations are usually scheduled according to the time scale of days, weeks, months, and years. Therefore, delay propagation between airports will also show timing patterns at various time scales. To take advantage of the temporal dependence of delay propagation, we construct causality graph at time $t$  from four independent time periods: $C^t = \left\{C^{t,year},C^{t,month},C^{t,week},C^{t,day}\right\}$. 

The causal inference is based on the stationarity of the sequence, and the flight delay sequence has strong non-stationarity. Therefore, we have implemented a differential trend method to construct a new sequence $\overline{Y^t}$ by calculating the differences between the uniform time interval. Subsequently, the Granger causality test measures the causal relationship between delay sequences of two airports, which is ideal for identifying whether one time series causes changes in another time series. 
For example, taking the delay state sequences $\overline{Y^t_a},\overline{Y^t_b}$ of airport $a$ and $b$ as input, the Granger causality test applied an unrestricted regression of $\overline{Y^t_a}$ with $\overline{Y_a^{t-l}}$ and $\overline{Y_b^{t-l}}$:
\begin{equation}
\begin{aligned}
& {\overline{Y^t_a}}=\alpha_0+\sum_{i=1}^l \alpha_i {\overline{Y_a^{t-l}}}+\sum_{j=1}^l \beta_j {\overline{Y_b^{t-l}}}+u_{1 t}
\end{aligned}
\end{equation}
where $u_{1t}$ is assumed to be uncorrelated white noise error items, and the residual sum of squares is recorded as $RSS_u$. 
Second, a restricted regression equation of $\overline{Y^t_a}$ with $\overline{Y_a^{t-l}}$ is established:
\begin{equation}
\begin{aligned}
& {\overline{Y^t_a}}=\alpha_0+\sum_{i=1}^l \alpha_i {\overline{Y_a^{t-l}}}+u_{1 t}
\end{aligned}
\end{equation}
where $RSS_r$ represents the residual sum of squares. Finally, F-statistic is adopted to compute the effect of $\sum_{j=1}^l \beta_j \overline{Y_b^{t-l}}$ on the regression through testing how significant the difference between $RSS_u$ and $RSS_r$, which is recorded by p-value~\cite{CHENshenwen}.  If the p-value is less than the chosen significance level $\sigma$(5\% by default), there is a causal relationship between airport $a$ and airport $b$, $C_{a,b}^t=1$; otherwise, $C_{a,b}^ t=0$. In summary, the Granger causality test is performed on each pair of airports, and a set $C^t$ of historical delay causality graphs in different time periods is obtained.\

\subsection{Self-Causal Correction Module} 
The ATS operates cyclically, and the changing trends in each cycle are complex, making the delay sequence highly non-stationary. 
Multiple differencing is a common method to eliminate the non-stationarity of a sequence, but this method is likely to destroy the time series information of the sequence itself, resulting in information loss, which in turn leads to deviations in calculating causal values.  Therefore, directly adopting the existing causal inference method for flight delay prediction may become less effective, and how to obtain accurate inter-airport causal association is a difficulty.

To obtain a more accurate causal relationship between airports, we designed a \textit{self-causality correction module} based on two contrast corrections to more effectively discerns the intensity of delay propagation at different moments. Specifically, we aggregate the node information of the causality graph $C^t$ calculated by Granger test with delay feature vector $X^t \in R^{B\times N \times D}$ at the current time $t$ and the previous hidden layer state $H^{t-1}\in R^{B\times N \times W}$ ($W$ is the dimension of the hidden layer and $B$ is batch size), and outputs $HC^t$:
\begin{equation}
HC^{t}=\alpha\left(X^t \| H^{t-1}\right)+\beta \Theta_{\star G}\left(X^t \| H^{t-1}, C^{t}\right)
\end{equation}
where ${\star G}$ is GCN operation, $\Theta$ represents the learnable parameters, $\alpha$ and $\beta$ are adjustable parameters, $\|$ represents the concatenate operation. Then, $HC^{t}$ with two different transformation matrices $E_1,E_2$ to obtain comparison matrices $\rho_1$ and $\rho_2$:
\begin{equation}
\begin{aligned}
& \rho_1=\mathit{tanh} \left(F C\left(H C^t\right) \odot E_1\right) \\
& \rho_2=\mathit{tanh} \left(F C\left(H C^t\right) \odot E_2\right)
\end{aligned}
\end{equation}
where $\odot$ denotes the Hadamard product, $FC(\cdot)$ is linear layer, and $tanh(\cdot)$ is the activation function. Finally, the causal graph is corrected by calculating the similarity of the comparison matrix~\cite{wu2020connecting}, and the correction mask $CM^{t}$ is obtained. If the similarity of the comparison matrix is high, it indicates that the causal graph is accurate and does not need to be greatly revised; otherwise, a certain correction is required:
\begin{equation}
C M^t=\mathit{Relu}\left(\mathit{tanh} \left(\rho_1 \cdot T\left(\rho_2\right)-\rho_2 \cdot T\left(\rho_1\right)\right)\right)
\end{equation}
where $tanh(\cdot)$ and $Relu(\cdot)$ are activation function, $T(\cdot)$ means transpose. Thus, we get the correction mask ${CM^{t}}=\{{CM^{t,year}}, CM^{t,month},CM^{t,week}, CM^{t,day} \in R^{B \times N \times N}\}$, which is a collection of causality graph corrections under four independent time periods. By linearly combining the correction mask with the causality graph, the self-corrected causal graph ${CA^t}$ is finally obtained. In general, the \textit{self-causal correction module} aggregates the information in the undifferentiated delay sequence through GCN, which can be regarded as a supplement to the missing causal value information. At the same time, the similarity calculations after two separate corrections are not completely random adjustments, but are also part of the accurate support for training data.
\begin{figure}[!t]
  \centering
  \includegraphics[width=3.4in]{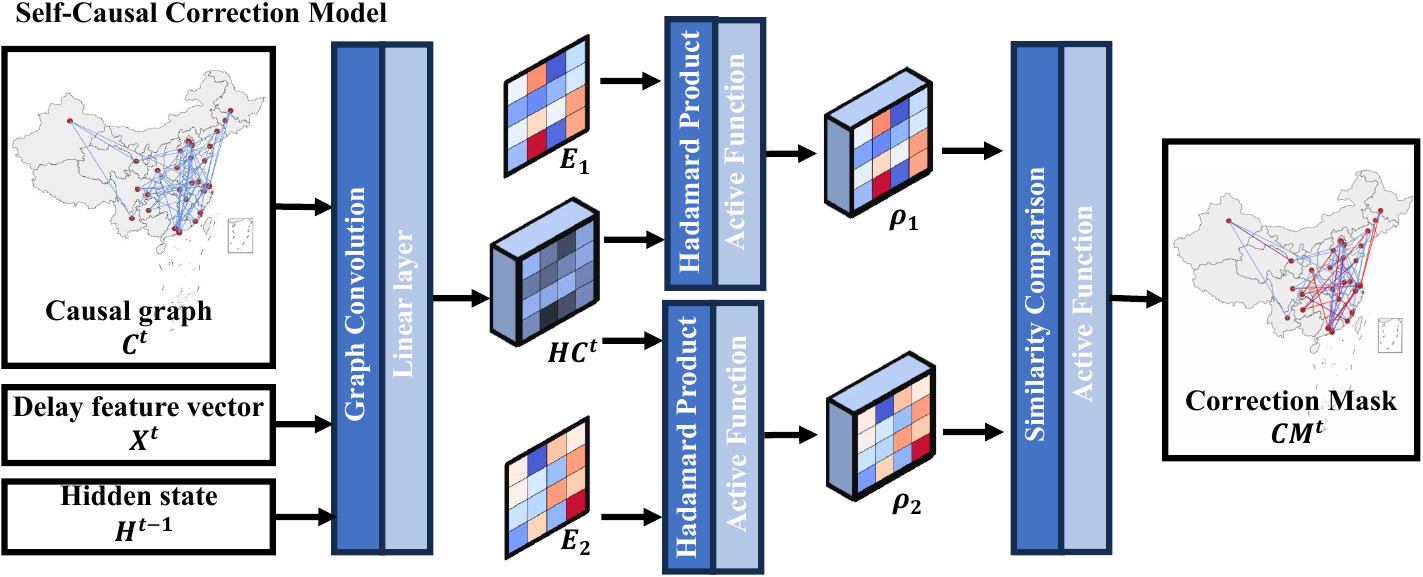}
\caption{Self-Causal Correction Module. This figure shows the specific process of the Self-Causal Correction Module. First, the causal graph is aggregated through GCN to aggregate node features. The output $HC^t$ will perform matrix operations with two transformation matrices $E_1$ and $E_2$ to obtain comparison matrices $\rho_1$ and $\rho_2$. By calculating the similarity between the comparison matrices, the degree of correction needed for the causal graph can be determined, and a correction mask $CM^{t}$ is constructed accordingly. The blue edges in the correction mask represent the original causal relationships, and the red edges represent the causal relationships that need to be corrected.}
\label{fig_2}
\end{figure}

\subsection{Spatial Dependence Modeling} 
When a flight delay spreads to the airport, due to differences in airport capacity, operation management and other structural characteristics, the efficiency of handling the delay is different, so the delay may be absorbed quickly or further expanded. Therefore, there is heterogeneity between airports in the air traffic system, which means that different airports are affected by delay propagation effects to varying degrees. Traditional GCN modules are incapable of extracting the heterogeneity of airports from the causal graph. Moreover, simply integrating them into existing STGNNs may not effectively represent the spatial correlations of flight delays. Thus, addressing the characteristics of each airport is a crucial step in accurately extracting spatial correlation information.

Accordingly, we proposed a graph information extraction method that integrates airport heterogeneity and causality to jointly characterize delay spatial dependence. Specifically, we aggregate the self-corrected causal graph $CA^{t_i}(i={year,month,week,day})$ and geographical graph $A$ respectively based on \textit{K}-hop GCN~\cite{DGCRN} to to obtain causal fusion graph $FC^t$ and geographic fusion graph $FA^t$ at time $t$:
\begin{equation}
\begin{gathered}
F C^t=\omega_{i n} h_{i n}+\sum_i \omega_i \Theta_{\star G}\left(h^{(k-1)}, \widetilde{CA}^{t_i}\right) \\
F A^t=\omega_A \Theta_{\star G}\left(h^{(k-1)}, \tilde{A}\right) \\
\widetilde{C A}^{t_i}=\widetilde{D}^{t^{-1}}\left(C A^{t_i}+I\right) \\
\widetilde{D}_{:, m, m}^t=1+\sum_n C A_{:, m, n}^{t_i} \\
\tilde{A}=\widetilde{D}^{-1} A, \widetilde{D}_{m, m}=\sum_n A_{m, n}
\end{gathered}
\end{equation}
where $\omega_{i n}$, $\omega_i$ and $\omega_A$ are trainable parameters, $h^{(k-1)}$ is the state of the $k-1$ hidden layer of \textit{K}-hop GCN, $h^{(0)}=h_{i n}$. The causal fusion graph $FC^t$ is then further fused with the and geographic fusion graph $FA^t$ through an adaptive weight embedding:
\begin{equation}
h^{(k)}=F I T_1 \odot F C^t+F I T_2 \odot F A^t
\end{equation}
$FIT_1$ and $FIT_2$ are trainable adaptive weighting matrices, which can adjust the weights between different graphs and iterate continuously in \textit{K}-hop GCN to achieve efficient heterogeneity fusion, and get the final output $h^{out}=\sum_k W^{(k)} h^{(k)}$ . In practical applications, we often use two-way \textit{K}-hop GCN, which is $h_{o}=\Theta_{1K \star G}\left(h_{i n}, C A^t, A\right) + \Theta_{2K \star G}\left(h_{i n}, T(CA^t), T(A)\right)$. The delay heterogeneity fusion process is simply expressed as: $h_{o}=\Theta_{K \star G}\left(h_{i n}, C A^t, A\right)$.


\subsection{Temporal Dependence Modeling} 
Considering the long-run dependencies in the airport delay prediction problem~\cite{CHENshenwen}, we design a novel time series prediction unit, named Long-Gate Recurrent Unit (LGRU), based on the Encoder-Decoder architecture. This unit can preserve long-term memory and outperform the traditional time series model in predicting delays. Specifically, in the Encoder part, the delayed feature vector $X^t$ and the historical hidden layer state vector $H^{t-1}$ at the previous moment are sequentially input into the LGRU module:
\begin{equation}
\begin{gathered}
re^{(t)}=\sigma\left[\Theta_{K \star G}\left(X_t \| H^{t-1}, C A^t, A\right)\right] \\
z^{(t)}=\sigma\left[\Theta_{K \star G}\left(X_t \| H^{t-1}, C A^t, A\right)\right] \\
o^{(t)}=\sigma\left[\Theta_{K \star G}\left(X_t \| H^{t-1}, C A^t, A\right)\right] \\
lm^{(t)}=\mathit{tanh} \left[\Theta_{K \star G}\left(X_t \| H^{t-1}, C A^t, A\right)\right]
\end{gathered}
\end{equation}
where $re^{(t)}$ is the reset gate, $z^{(t)}$ is the update gate, $o^{(t)}$ is the output gate, $lm^{(t)}$ is a candidate long-term vector; in LGRU, we remove the long-term information reset to retain the long-term information, extract time dependence and space dependence at the same time after merging with historical information, and finally output the hidden layer $H^t$ at moment $t$:
\begin{equation}
H^t=o^{(t)} \odot\left[\mathit{tanh} \left(z^{(t)} \odot H^{t-1}+re^{(t)} \odot lm^{(t)}\right)\right]
\end{equation}
Decoder first receives the hidden layer of the last output of the Encoder as the initial hidden layer input, and still uses the LGRU to iteratively decode. The slight difference is that at time $t+m$, the historical delay causality graph set $C^{t+m-T}$ is used as the input of the causal graph to construct the causal correction matrix. At the same time, since the delay features at time $t+m$ are unknown, the input is replaced by an all-zero matrix with the same size as $X^t$. Subsequently, the linear layer transforms the final output of the Decoder into prediction results. 

\section{Result}
\subsection{Experiments and Analysis}
This article utilizes a dataset provided by the Civil Aviation Administration of China (CAAC), covering the delays at 224 airports in China from April 1, 2018 to October 31, 2018. Considering flights tend to be concentrated in a few airports, 224 civil airports are ranked according to their traffic flow, and the top-74 busiest airports are used to evaluate the performance of the proposed model. These airports incorporate information from approximately 2.19 million scheduled flights, accounting for over 90\% of the total air traffic volume in China during the period. Considering that some airports have extreme values with large deviations in non-operating times, we removed 4.26\% of outliers before constructing the dataset. \
Flight delay data is logged on an hourly basis, resulting in 5,136 samples for each airport over the span of 214 days. 70\% of the data is used as a training set, 15\% is used as a validation set, and the remaining 15\% is used as a test set. To eliminate the scale difference in the data, Z-Score standardization is performed on all the original data:
\begin{equation}
z_i=\frac{x_i-\mu}{\sigma}
\end{equation}
where $x_i$ is the original data sample, $\mu$ is the mean of the data, and $\sigma$ is the standard deviation of the data. Consistent with the existing research, we predict the average flight delay of each airport every hour, and the prediction time limit is 1 to 3 hours.

\subsection{Baseline Methodologies and Evaluation Metrics}
We compared the proposed CausalNet with single-airport scenario methods and multi-airport scenario methods.  Two representative methodologies that have been adopted to predict flight delay in the single airport scenario which are SVR and LSTM. In the multi-airport scenario, we compared several well-known STGNNs, including GCGRU, DCRNN, STGCN and the state-of-the-art DGCRN. Each method is applied to predict flight delays and evaluate its accuracy. Furthermore, our comparison is extended to MSTAGCN, a STGNN framework specifically used in flight delay prediction, which is widely recognized due to its cutting-edge and accuracy. 

Through the above baseline comparison, we can determine how advanced CausalNet is, as well as the strengths and limitations of each model, providing valuable insights into its practical applications and potential improvements.

\begin{itemize}
\item \textbf{\textit{SVR}}~\cite{SVR}: Support Vector Regression is a method of using SVM (Support Vector Machine) to fit curves and do regression analysis. In this paper, the polynomial kernel function (RBF) is used to map to the feature space and then perform regression, where the error term penalty coefficient C=0.2.
\item \textbf{\textit{LSTM}}~\cite{LSTM}:  Short-term memory network is an improved RNN, which can use time series model to analyze input, and can realize the effective use of long-distance time series information. The initial learning rate is 1e-3 with a decay rate of 0.6 after every 10 epochs, and the batch size is 64.	
\item \textbf{\textit{STGCN}}~\cite{STGCN}: Spatial-temporal Graph Convolutional Network is a typical STGNN. The channels of three layers in ST-Conv block are 64, 16, 64 respectively. Both the graph convolution kernel size and temporal convolution kernel size are set to 3. The initial learning rate is 1e-4 with a decay rate of 0.6 after every 5 epochs. The total number of epochs is set to 50, and the batch size is 64.
\item \textbf{\textit{DCRNN}}~\cite{DCRNN,ITSM_new2}: Diffusion Convolutional RNN, which captures spatial dependence through the operation of diffusion convolution. Both encoder and decoder contain two recurrent layers and there are 64 units in each recurrent layer. The initial learning rate is 1e-4 with a decay rate of 0.9 after every 10 epochs, and the filter size is 3.
\item \textbf{\textit{GCGRU}}~\cite{GCGRU}: Graph Convolutional GRU replaces the matrix multiplication in gated recurrent unit with the graph convolution. The GCGRU applied in this paper contains a GCGRU layer with a normalized Laplacian matrix for the traffic flow graph. The initial learning rate is 1e-3 with a decay rate of 0.9 after every 10 epochs, and the batch size is 64.
\item \textbf{\textit{DGCRN}}~\cite{DGCRN}:Dynamic Graph Convolutional Recurrent Network is a GNN and RNN-based model where the dynamic adjacency matrix is designed to be generated from a hyper-network step by step in synchronization with the iteration of the RNN. The subgraph size is 20, the dimension of node embedding is 40, the hidden state dimension of RNN is 64, and the hidden layer dimension of hypergraph neural network is 32. The initial learning rate is 1e-3 and batch size is 64.
\item \textbf{\textit{MSTAGCN}}~\cite{Cai}: Multiscale Spatial-Temporal Adaptive Graph Convolutional Neural Network for the network-wide multi-step-ahead flight delay prediction problem. The graph kernel size is set as 5, The time-window is set to 24 and batch size is 64. The initial learning rate is 1e-4, with a decay rate of 0.6 after every 5 epochs.

\end{itemize}
\subsection{Experimental Setup}
The proposed model is implemented by Pytorch 1.11 on a virtual workstation with a 12 GB Nvidia GeForce GTX 1080Ti GPU. The delay feature used in this article is the corresponding flight delay time, and the lag order for causal inference $l=2$. The experiment is repeated 5 times, and the average value of the assessment metrics is reported. Utilizing the Adam optimizer for 150 epochs with a 64-batch size, and the model is trained by reducing the mean absolute error. The use of early stopping helps prevent overfitting. After every 5 epochs, the learning rate decays by 0.6, with an initial learning rate of 1e-4. The size of the hidden state is set to 64, while the node embeddings have a dimension of 40. In the initial step, vectors of 1 are specified for $\omega_{i n}$,  $\omega_{i}$,  $\omega_{A}$, $FIT_1$ and $FIT_2$. The learnable parameter weights or bias are initialized using Kaiming methodology~\cite{he2015delving} and the user-defined parameters are determined through the grid-search methodology.

\subsection{Evaluation Metrics}
To conduct fair comparisons, two commonly used performance indexes are employed to evaluate the performance of the models. They are Mean Absolute Error (MAE), and Root Mean Squared Error (RMSE), which can be defined as:
\begin{equation}
\begin{aligned}
M A E & =\frac{1}{k} \sum_{i=1}^k\left|\hat{y}_i-y_i\right|, \\
R M S E & =\sqrt{\frac{1}{k} \sum_{i=1}^k\left(\hat{y}_i-y_i\right)^2}
\end{aligned}
\end{equation}
where $k$ is the number of testing samples. $y_i$ and $\hat{y}_i$ denote the actual observations and model predictions of flight delay, respectively. MAE measures the average absolute error between model predictions and actual observations. It is simple and intuitive but does not pay attention to outliers with large deviations. RMSE measures the square root of the average squared error between model predictions and actual observations, and is more sensitive to outliers or values with large errors. Measuring MAE and RMSE simultaneously helps us evaluate the accuracy of prediction results from multiple perspectives.

\subsection{Results and Analysis}
\subsubsection{Performance Comparison}
To present the results in a statistically significant way, we carry out 60 seeded runs and use the mean of the MAE and RMSE to evaluate the performance of various models. The prediction results for the next 1-3 hours are reported in Table $\text {I}$. \textbf{CausalNet} outperforms the baselines across all time frames, exhibiting the lowest MAE and RMSE. This highlights the superiority of our model in predicting flight delays.

Fig.3 illustrates the improvement ratio of CausalNet compared to the baselines. 
It appears that the performances (MAE/RMSE) of CausalNet are average 15.033\%/11.508\% lower than those of baselines on short term predictions (next 1h) and 14.600\%/4.426\% lower on long term predictions (next 2-3h). Additionally, it is noted that the performances of SVR and LSTM are the worst among all methods, as they are proposed for the single-airport scenario without considering the spatial information of flight delays among airports. GCGRU, DCRNN, STGCN, DGCRN, MSTAGCN outperform others since they can extract both spatial features and temporal features.

Specially, two state-of-the-art methods DGCRN and MSTAGCN work well because they are not based solely on routes or geographical distances when extracting flight delay correlations between airports, but are adjusted based on observational data. Compared with these two methods, CausalNet still has an improvement rate of 4.141\%/4.483\%, indicating that the proposed self-corrective causal module and spatial dependece module can more accurately capture and use the delay propagation correlation between airports.

The following sections will dive deeper into the contribution of each design element to delay prediction through a comprehensive ablation study. It is noted that although the improvement ratio of our model decrease with longer look-ahead times, it still maintains an advantage of 16.8464\% of MAE and 4.5630\% of RMSE for the next 3 hours. The primary reason for this degradation may be that the causal graph mainly captures the delay correlation at the current moment, leading to better short-term prediction performance. Due to the dynamic nature of flight delays, the causal graph at the current moment may have limited impact on long-term prediction. As look ahead of time increases, the effect of spatial information at the current time gradually weakens, resulting in performance degradation. 

\begin{figure*}[t]
  \centering
  \includegraphics[width=\linewidth]{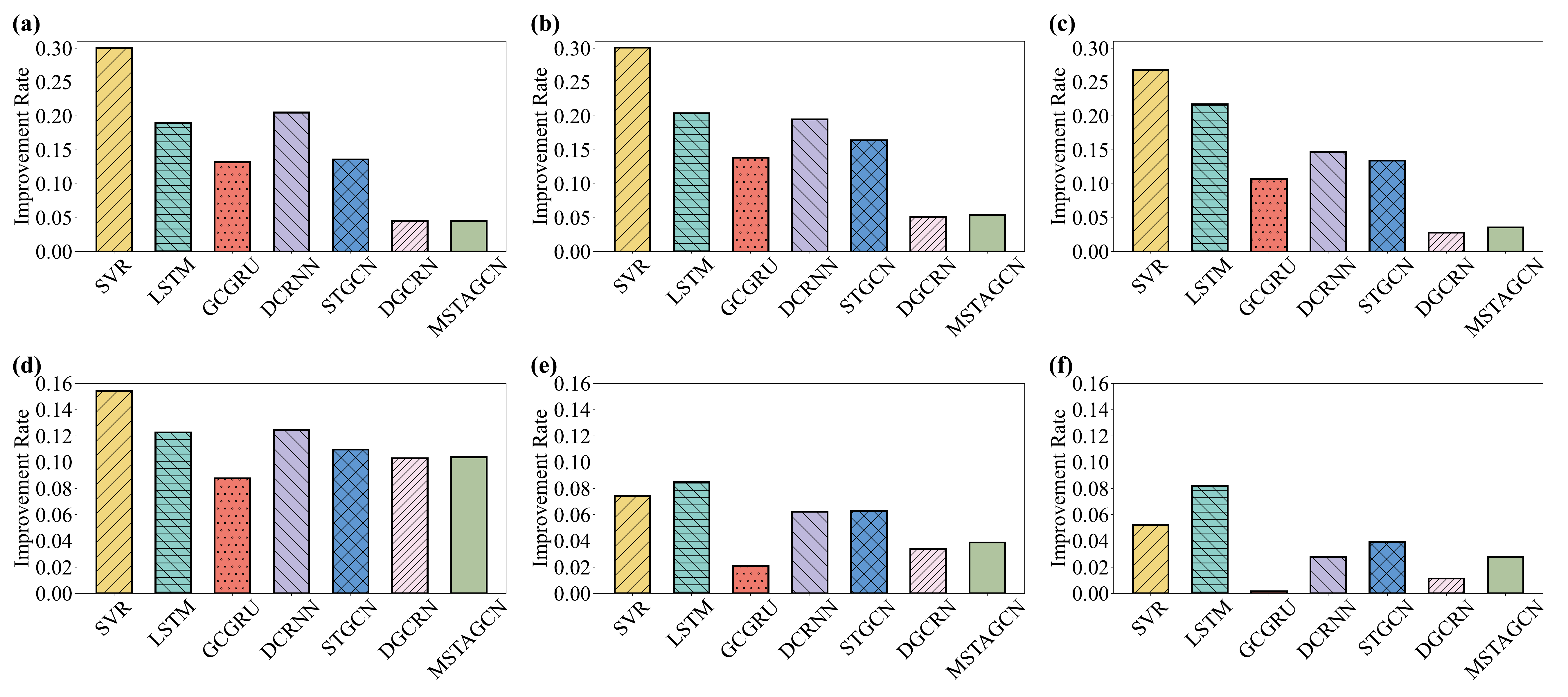}
  \caption{Compared with the improvement rates of different algorithms in MAE and RMSE. (a)(b)(c) shows the proportion of different algorithms' MAE exceeding our model within 1 to 3 hours, and (d)(e)(f) shows the proportion of different algorithms' RMSE exceeding our model within 1 to 3 hours.}
  \label{fig_3}
\end{figure*}

\begin{figure*}[t]
  \centering
  \includegraphics[width=\linewidth]{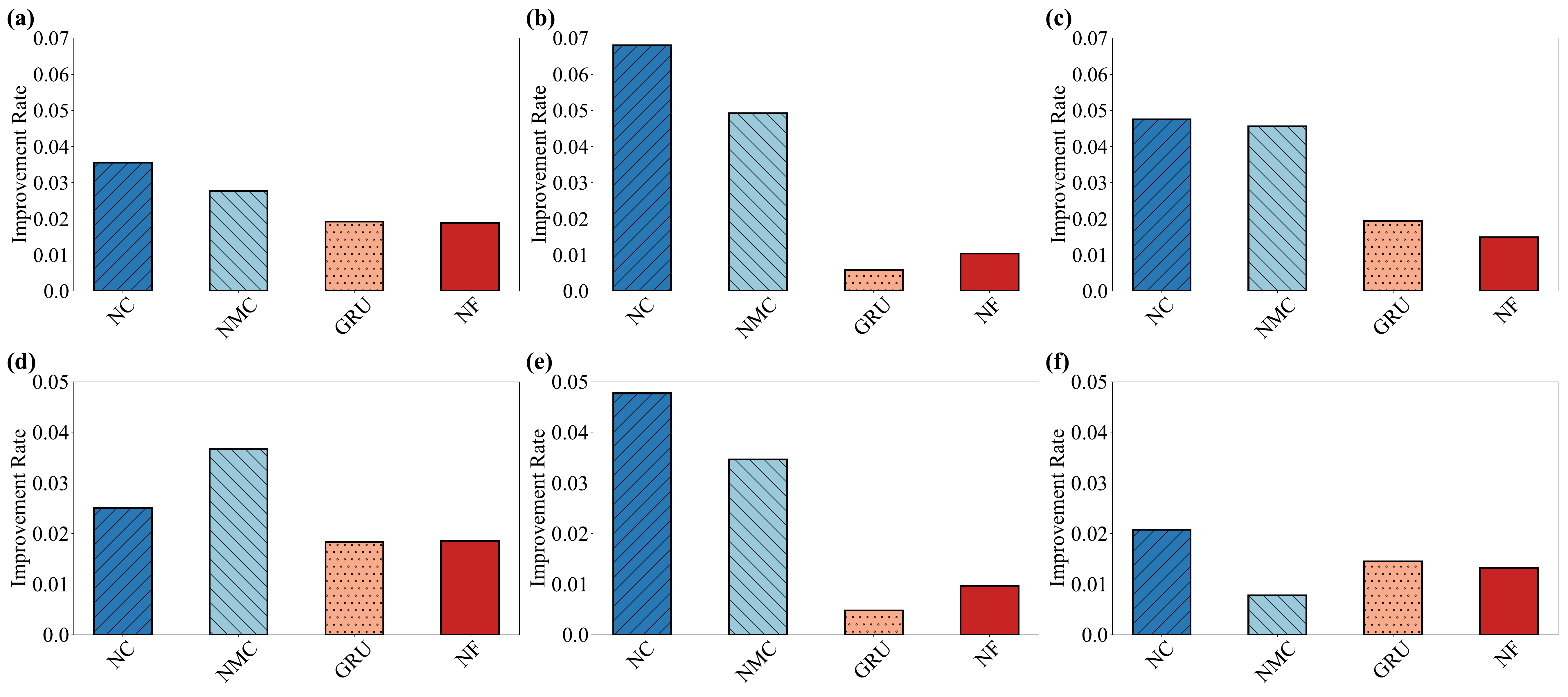}
  \caption{Performance of the ablation experimental model on MAE and RMSE. (a)(b)(c) shows the proportion of MAE of different CausalNet removing some components that is higher than the original model within 1 to 3 hours. (d)(e)(f) shows the proportion above RMSE from 1 to 3 hours.}
  \label{fig_4}
\end{figure*}


\begin{table*}[h]
    \caption{\textbf{Performance comparison of baselines}}
    \centering
    \begin{tabular}{cccccccccc}
    \toprule
   \textbf{Look ahead of time}& \textbf{Evaluate  Metric}& \textbf{SVR} & \textbf{LSTM} & \textbf{GCGRU} & \textbf{DCRNN} & \textbf{STGCN} & \textbf{DGCRN} & \textbf{MSTAGCN}& \textbf{CausalNet}\\
    \midrule
    \multirow{2}{*}{1h}& MAE & 8.024 & 6.932 & 6.471 & 7.067 & 6.501 & 5.883 & 5.884& \textbf{5.618}\\
     & RMSE & 10.992 &10.595  & 10.189 & 10.618 & 10.439 & 10.363 & 10.371& \textbf{9.295}\\
    \midrule
    \multirow{2}{*}{2h}& MAE & 8.286 & 7.277 & 6.724 & 7.197 & 6.932 & 6.106 & 6.123& \textbf{5.794}\\
     & RMSE & 11.019 &11.150  & 10.419 & 10.877 & 10.883 & 10.558 & 10.613& \textbf{10.201}\\
    \midrule
    \multirow{2}{*}{3h}& MAE & 8.236 & 7.705  & 6.752 & 7.071 & 6.963 & 6.204 & 6.252& \textbf{6.030}\\
    & RMSE & 11.063 & 11.424  & 10.505 & 10.788 & 10.913 & 10.608 & 10.787& \textbf{10.488}\\
    \bottomrule
    \end{tabular}
\end{table*}

\begin{table*}[h]
    \caption{\textbf{Ablation Study}}
    \centering
    \begin{tabular}{ccccccc}
    \toprule
    \textbf{Look ahead of time}& \textbf{Evaluate Metric}& \textbf{CausalNet-NC}& \textbf{CausalNet-NMC}& \textbf{CausalNet-GRU}& \textbf{CausalNet-NF}& \textbf{CausalNet}\\
    \midrule
    \multirow{2}{*}{1h}& MAE & 5.825 & 5.778 & 5.728 & 5.726 & \textbf{5.618}\\
    & RMSE & 9.534 & 9.649 & 9.468 & 9.471 & \textbf{9.295}\\
    \midrule
    \multirow{2}{*}{2h}& MAE & 6.217 & 6.094 & 5.828 & 5.855 & \textbf{5.794}\\
    & RMSE & 10.712 & 10.567 & 10.250 & 10.300  & \textbf{10.201}\\
    \midrule
    \multirow{2}{*}{3h}& MAE & 6.331 & 6.318 & 6.149 & 6.121 & \textbf{6.030}\\
    & RMSE & 10.710 & 10.570 & 10.642 & 10.628 & \textbf{10.488}\\
    \bottomrule
    \end{tabular}
\end{table*}

\subsubsection{Ablation Study}
We conducted ablation experiments to engage in a more in-depth analysis of the designed modules, with a focus on evaluating the performance of models from which specific components have been removed. In these experiments, four distinct models were employed, each designated by a unique name: “CausalNet-NC”, “CausalNet-NMC”, “CausalNet-GRU”, and “CausalNet-NF”. The “CausalNet-NC” model is characterized by the removal of the \textit{granger causal graph construction module} and \textit{self-causal correction module}. In the “CausalNet-NMC”, the \textit{self-causal correction module} is eliminated and only the initial causal graph and aggregated node information are input to subsequent modules. For “CausalNet-GRU”, \textit{LGRU} is replaced by GRU. Moreover, “CausalNet-NF” features the removal of the \textit{adaptive weight matrix} in \textit{spatial dependence modeling}. Table $\text {II}$ comprehensively lists the performance metrics of each model. 

The models “CausalNet-NC”, “CausalNet-NMC”, “CausalNet-GRU”, and “CausalNet-NF” exhibit average MAE decline rates of 5.328\%/4.273\%/1.506\%/1.494\%, respectively, from 1 to 3 hours, with corresponding average RMSE decline rates of 3.232\%/2.728\%/1.274\%/1.399\%. Notably, “CausalNet-NC” consistently underperforms at all look ahead times, underscoring the significance of incorporating the causal graph for accurate flight delay time predictions. The delay causality graph effectively aids the model in capturing delay correlations more precisely.

As the prediction time prolongs, due to the inherent periodicity, trend and other non-stationary characteristics of flight delay data, the accuracy of delay causality extraction decreases. In this context, integrating the \textit{self-causal correction module} becomes increasingly important to improve performance.  At the 3-hour mark, the MAE of the CausalNet model with the \textit{self-causal correction module} improves by 0.2882 compared to “CausalNet-NMC”, while "CausalNet-NC" only reaches 6.3313 relying only on delayed causality graphs as input. This verifies the effectiveness of the causal correction module.

Furthermore, results related to “CausalNet-NF” demonstrate that implementing an \textit{adaptive weight matrix} in \textit{spatial dependence modeling} facilitates the efficient integration of causal correlations while considering airport heterogeneity characteristics. The poor performance of “CausalNet-GRU” proves that \textit{LGRU} can be more suitable for long temporal dependency extraction in flight delay prediction problems.

\subsection{Evaluation of the Self-Causal Correction Module}
Causal inference provides a data-driven approach for identifying relationships in flight delays between airports. Different from the correlation coefficients used in previous works, causal inference detects the dependences among airports rather than only the statistical association. The fundamental assumption of causal inference is that the used sequences are stationary. However, the evident periodicity and trends in flight operations render the delay time series a typical non-stationary series.  
Therefore, to accurately calculate causal graphs, a \textit{self-causal correction module} with trainable parameters is further designed to adaptively modify the elements of the causal graph.

Fig. 5 shows the correction results of delay causality graphs at different time scales. It appears that the causal graph requires the most correction on an annual scale and the least on a daily scale, indicating that with the accumulation of time series, the causal calculation deviation increases. To demonstration, we calculate the Euclidean distances between the corrected graphs and the original causality graphs. For graphs at year, month, week, and day, the distances are 223.579, 146.949, 122.581, and 100.874 respectively.
The above conclusion is consistent with reality. In actual air traffic systems, flight operation non-stationarity accumulates over time and becomes more obvious in long-term delay time series, leading to large deviations in causal calculations.

\begin{figure}[t]
 \centering
  \includegraphics[width=\linewidth]{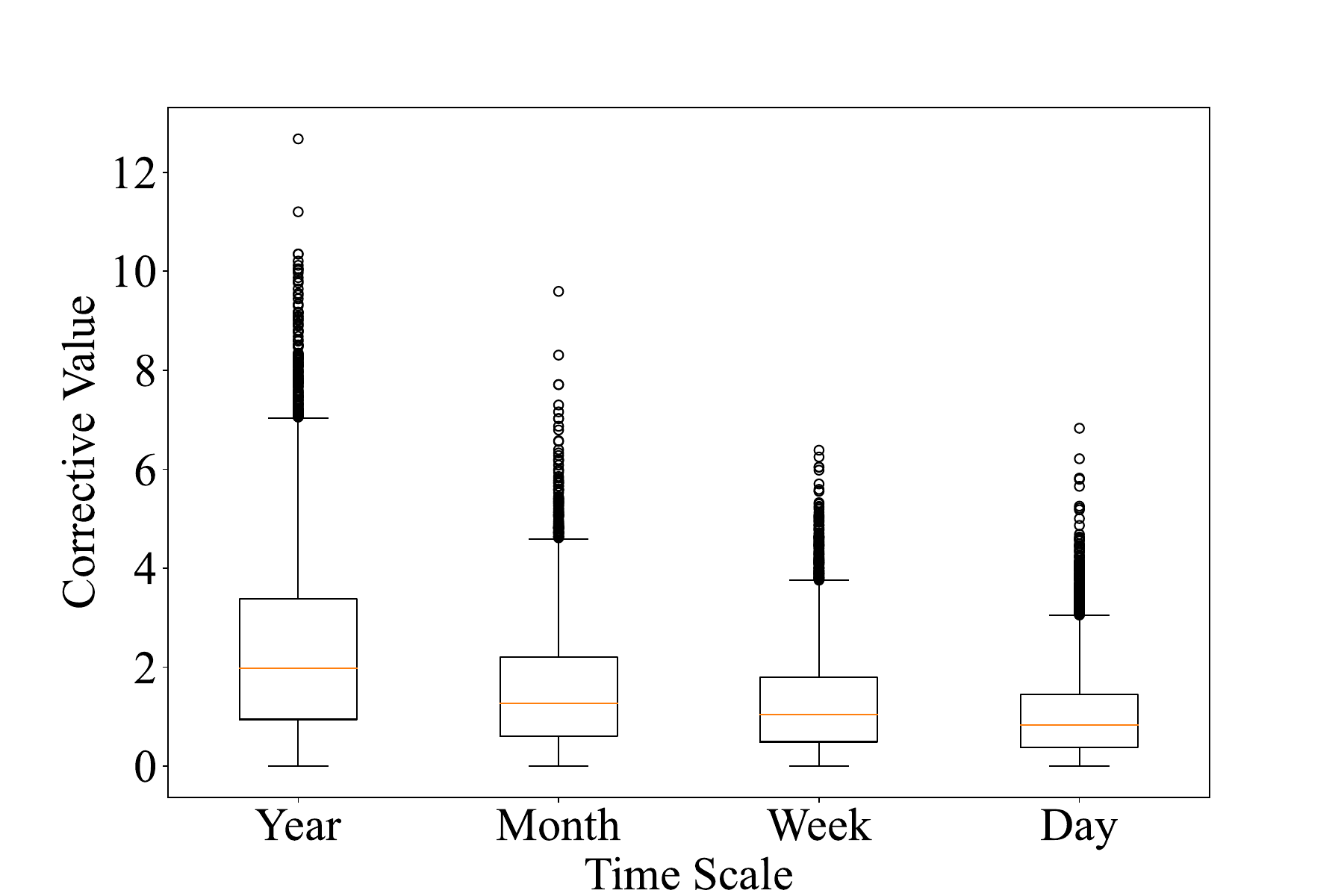}
\caption{Causal correction result. This figure shows the differences between the original causal graph and the corrected causal graph from four scales: year, month, week, and day. }
\label{fig_5}
\end{figure}

\subsection{Analysis of Adaptive Parameter Matrix}
\begin{figure}[t]
  \centering
  \includegraphics[width=\linewidth]{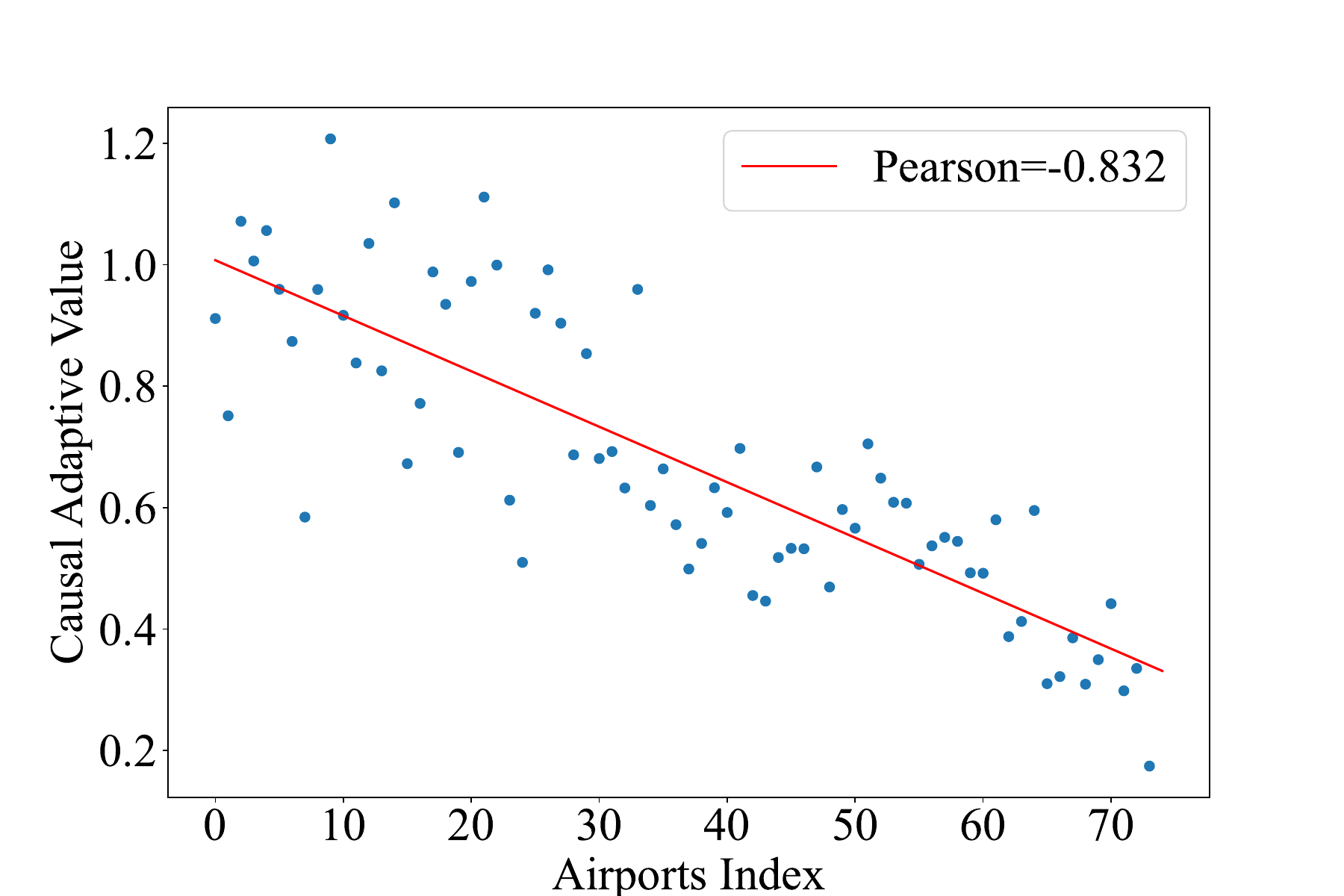}
  \caption{The change in adaptive value associated with small to large airport delays.}
  \label{fig_6}
\end{figure}
The parameters within the adaptive weighting matrix are dynamically updated during the training process with input data, thereby reflecting the dependency of different airports on the delay propagation from other airports. A higher weight assigned to a parameter indicates that delays at the corresponding airport are significantly influenced by delays at other airports. Conversely, a lower weight suggests that the airport is less susceptible to delays at other airports.

Fig. 6 displays the updated parameters for each airport, revealing a clear negative correlation between the weight of parameters and the airport's traffic ranking. This implies that larger airports are less susceptible to external influences and experience delays primarily due to their own characteristics. In contrast, smaller airports are more vulnerable to delay propagation from other airports.


This phenomenon can be attributed to the Civil Aviation Administration of China has implemented a no take-off limit regulation for larger airports. Flights departing from these airports are not subject to traffic flow management initiatives~\cite{DU2018}. Furthermore, as these major airports often operate close to their capacity limits, air traffic controllers frequently implement traffic flow management strategies. Such plans involve delaying incoming flights from upstream airports to align the traffic demand at major hubs with their operational capacity, thereby ensuring smoother and more efficient handling of air traffic.



\section{Conclusion}
This paper proposes CausalNet, a spatial-temporal graph neural network with a novel self-corrective causal inference for accurate flight delay prediction.
A real-world dataset encompassing the 74 busiest airports in China is utilized in the experiments. The results compellingly show that the performances (MAE/RMSE) of CausalNet are average 15.033\%/11.508\% lower than those of baselines on short term predictions (next 1h) and 14.600\%/4.426\% lower on long term predictions (next 2-3h). Subsequent ablation analyses further underline the effectiveness of the causal graph and our innovative self-correcting causal module for predicting flight delays. 
By further comparing the differences in causal corrections at different time scales, we found that flight operation non-stationarity accumulates over time, which is consistent with the actual situation. 
Moreover, by exploring the impact of delay propagation on delays at different airports, we found that large airports are less susceptible to the impact. These provide insights into actual flight operations.

Exploring multi-airport delay prediction based on deep spatio-temporal graph networks can be further extended. We note that the impact of flight delay propagation may be foreshadowed several hours in advance, so delays can take into account not only information from the current time, but also from previous times. How to use graph convolution to aggregate information from previous moments has become a current difficulty. Furthermore, the delay prediction problem can focus on accurate prediction of individual flights, which may be more effective for aviation network operations.

\ifCLASSOPTIONcaptionsoff
  \newpage
\fi

\vfill

\end{document}